\documentclass[aps,pra,twocolumn]{revtex4-2}  
\usepackage{amsmath,amssymb}  
\usepackage{graphicx}         
\usepackage{booktabs}
\usepackage{natbib}
\usepackage[colorlinks=true, linkcolor=blue, citecolor=blue, urlcolor=blue]{hyperref} 
\usepackage{tabularx}
\usepackage{bm}
\usepackage{braket}
\usepackage{orcidlink}

\def\HIM{Helmholtz Institute Mainz, 55099 Mainz, Germany}
\def\GSI{GSI Helmholtzzentrum für Schwerionenforschung GmbH, 64291 Darmstadt, Germany}
\def\JGU{Johannes Gutenberg University, Mainz 55128, Germany}

\def\UC{Department of Physics, University of California at Berkeley, Berkeley, California 94720-7300, USA}
\def\CSUEB{Department of Physics, California State University -- East Bay, Hayward, California 94542-3084, USA}

\def\PNPI{Petersburg Nuclear Physics Institute of NRC ``Kurchatov Institute'', Gatchina 188300, Russia}
\def\ETU{St.Petersburg Electrotechnical University LETI, Prof. Popov Str. 5, 197376 St. Petersburg, Russia}

\begin{document}

\title{Constraints on exotic interactions from scalar spin-spin coupling in tritium deuteride (DT)}

\author{Lei Cong$^{1,2,3}$\orcidlink{0000-0003-0002-1840},
Derek~F.~Jackson Kimball$^{4}$\orcidlink{0000-0003-2479-6034}, 
Mikhail G.~Kozlov$^{5,6}$\orcidlink{0000-0002-7751-6553},
and Dmitry Budker$^{1,2,3,7,*}$\orcidlink{0000-0002-7356-4814}}
\address{
$^{1}$ \HIM\\
$^{2}$ \GSI\\
$^{3}$ \JGU\\
$^{4}$ \CSUEB\\
$^{5}$ \PNPI\\
$^{6}$ \ETU\\
$^{7}$ \UC\\
* budker@uni-mainz.de
 }

\begin{abstract}
A comparison of theoretical and experimental values of the scalar spin-spin interaction ($J$-coupling) in tritium deuteride molecules yield constraints for nucleon-nucleon exotic interactions of the dimensionless coupling strengths $g_Vg_V$, $g_Ag_A$ and $g_pg_p$, corresponding to the exchange of an vector, axial-vector, and pseudoscalar (axionlike) boson. 
The couplings between proton ($p$) and nucleon ($N$), denoted by $g_V^p g_V^N$, $g_p^p g_p^N$ are constrained to be less than $1.4 \times 10^{-6}$ and $2.7\times 10^{-6}$, respectively, for boson masses around 5\,keV. The coupling constant $g_A^p g_A^N$ is constrained to be less than $1.0 \times 10^{-18}$ for boson masses $\leq 100$\,eV. 
It is noteworthy that this study represents the first instance in which constraints on $g_V g_V$ have been established through the analysis of the potential term $V_2 + V_3$ for both tritium deuteride and hydrogen deuteride molecules.

\end{abstract}

\maketitle

\section{Introduction}

Exotic spin-dependent interactions have gained significant traction \cite{safronova_search_2018}, primarily as potential signatures of new bosons. Possible candidates, such as the axion \cite{dine_simple_1981,zhitnitsky_possible_1980, kim_weak-interaction_1979, shifman_can_1980, weinberg_new_1978, wilczek_problem_1978} and paraphoton \cite{holdom_two_1986,appelquist_nonexotic_2003,dobrescu_massless_2005}, offer explanations for several enduring puzzles in modern physics, including the origins of dark matter \cite{preskill_cosmology_1983,abbott_cosmological_1983,dine_not-so-harmless_1983,jackson_kimball_search_2023} and dark energy, thereby driving extensive experimental searches of spin-dependent interactions \cite{moody_new_1984,dobrescu_spin-dependent_2006,fadeev_revisiting_2019} across subatomic and astrophysical domains \cite{ficek_constraints_2018,fadeev_pseudovector_2022,stadnik_improved_2018,ji_constraints_2023,jiao_experimental_2021,kim_experimental_2018,hunter_using_2013,heckel_limits_2013}.

Early works analyzed experimental data and theoretical calculations to explore exotic interactions for simple molecules: dipole-dipole interaction of hydrogen (H$_2$) \cite{ramsey_tensor_1979} and spin-spin $J$-coupling for deuterated molecular hydrogen (HD) \cite{ledbetter_constraints_2013} respectively, and produced upper limits on the strengths of exotic interactions between protons  ($p$-$p$) and proton and nucleon ($p$-$N$). 

In this work, we utilize existing measurements and calculations of $J$-coupling for another isotopic variant of H$_2$ and HD, the deuterium tritide (DT) molecule, to explore spin-dependent exotic interactions. Specifically, we focus on the data from the latest experimental work by \citet{garbacz_indirect_2016} and the latest theory by \citet{puchalski_nuclear_2018}, which considered corrections ignored in the earlier theoretical work and brought the experiment and theory into agreement with each other.

The $J$-coupling interaction \cite{jazwinski_theoretical_2020} has the form $J\,\bm{I}_1\cdot \bm{I}_2$ (here, $\bm{I}_{1,2}$ are nuclear spin operators), and is 
present \cite{hahn_chemical_1951,gutowsky_nuclear_1953} due to second-order hyperfine interaction \cite{ramsey_interactions_1952}. The $J$-coupling interaction, in contrast to the magnetic dipole-dipole interaction, does not average out by molecular tumbling and can be observed in fluid samples.

Our goal is to constrain exotic spin dependent interaction between nucleons 
via several potential terms, including $(V_2+V_3)|_{VV}$ [see Eq.\,\eqref{gVgV_V23} and Ref.\,\cite{fadeev_revisiting_2019}] that has not been studied yet.
The strong motivation to explore exotic spin-dependent interactions in the context of beyond-standard-model scenarios is discussed in \cite{safronova_search_2018,cong_spin-dependent_2024}.

\section{DT $J$-coupling results}
While there are various theoretical and experimental studies of H$_2$ and HD \cite{helgaker_nmr_2012,garbacz_hd_2014,garbacz_indirect_2016,neronov_determination_2018,puchalski_nuclear_2018,jankowska_novel_2015,enevoldsen_correlated_1998,enevoldsen_correlated_1998,vahtras_indirect_1992,oddershede_nuclear_1988}, DT is explicitly considered in only a few references \cite{garbacz_indirect_2016,puchalski_nuclear_2018,aleksandrov_study_2011}. 

\citet{garbacz_indirect_2016} investigated the $J$-coupling constants for DT using gas-phase nuclear magnetic resonance (NMR) at 300\,K. The experimental values were measured and corrected for intermolecular interactions and other density-dependent effects  using zero-density extrapolation. By comparison with full configuration interaction (FCI) level theoretical calculations, \citet{garbacz_indirect_2016} revealed discrepancies, proportional to the inverse of the reduced mass of DT that may arise from nonadiabatic effects. 

\citet{puchalski_nuclear_2018} used a variational approach to calculate the $J$-couplings in hydrogen molecules, achieving precise results for DT at 300\,K using the adiabatic approximation. They also pointed out that the theoretical-experimental difference ($\sim$ -0.09\,Hz) was likely caused by nonadiabatic effects and could be corrected with finite nuclear mass corrections, estimated with a value of -0.10\,Hz. 

Table\,\ref{table1} summarizes the latest theoretical and experimental results as well as the difference ($\mu$) between theory and experiment, the combined uncertainty ($\sigma$), and use these to derive $\Delta J$, representing the maximum deviation. 
The calculations are based on the following equations: $\mu = \text{Theory} - \text{Expt}$\,, $\sigma = \sqrt{{\sigma_{\textrm{th}}}^2 + {\sigma_{\textrm{expt}}}^2}$\,, where $\sigma_{\textrm{th}}$ and $\sigma_{\textrm{expt}}$ are uncertainties for theory and experiment, respectively. $\Delta J$ is derived from the integral equation given by:
\begin{equation} I = \int_{-\Delta J}^{\Delta J} \frac{1}{\sqrt{2\pi}\sigma} e^{-\frac{(x-\mu)^2}{2\sigma^2}} dx =95\%\,,
\end{equation}
where $95\%$ confidence level is chosen to give constraints for exotic interaction between neucleons in this work.

\begin{table}[!h]
\centering
\caption{Comparison of most accurate 
theoretical and experimental values of 
the constant $J$ for
DT to calculate $\Delta J$. A high-precision value of $J=45.6506(9)$ was obtained in \cite{puchalski_nuclear_2018} in the adiabatic approximation.  The estimated value for nonadiabatic effects is $-0.10$\,Hz \cite{puchalski_nuclear_2018} but it has not been carefully evaluated yet. In our calculation, we add this value to the theoretical value with a 100\% uncertainty. Other theoretical
and experimental,
results are discussed in App.\,\ref{Existing_exp}. 
}\label{table1}
\begin{tabular}{lccc}
\hline
\hline
Parameter & $J$-coupling (300K) & References \\
\hline
Theory (Hz)& 45.55(10) &\cite{puchalski_nuclear_2018} \\
Expt (Hz)& 45.56(2) &\cite{garbacz_indirect_2016}\\
\hline
$\mu$ (Hz)& 0.01\\
$\sigma$ (Hz) & 0.102\\
\hline
$\Delta J$ (Hz) (95\%)& 0.201\\
\hline
\hline
\end{tabular}
\end{table}

\section{Constraints for Exotic Interactions}


The theoretical framework for spin-dependent exotic interactions was built step by step \cite{moody_new_1984,dobrescu_spin-dependent_2006,fadeev_revisiting_2019}.
High-precision NMR measurements can be sensitive to exotic forces \cite{ledbetter_constraints_2013}. For example, the strength of an exotic force can be constrained by comparing the experimental results with theoretical predictions, similarly to what is done in atomic spectroscopy \cite{ficek_constraints_2018}, with trapped ions \cite{kotler_measurement_2014}, or in parity-violation studies \cite{antypas_isotopic_2019,dzuba_probing_2017}. 
Here, we use the most recent experimental results to extend the analysis to DT, obtaining constraints for nucleon-nucleon interactions and extending the analysis to a range of spin-dependent interactions \cite{fadeev_revisiting_2019} $g_V g_V$, $g_A g_A$ and $g_p g_p$.

\subsection{Vector/vector interactions}\label{sec:gvgv}

Following the theoretical framework presented in Ref.\,\cite{fadeev_revisiting_2019} and \cite{cong_spin-dependent_2024}, vector/vector interactions have contributions from not only spin-independent $V_1$ potential term, but also from spin-dependent $(V_2+V_3)|_{VV}$ and $V_{4+5}|_{VV}$ potential terms. The $(V_2+V_3)|_{VV}$ term was not, to our knowledge, studied before. The only similar term is $(V_2+V_3)|_{AA}$, studied in \cite{fadeev_pseudovector_2022} which has a factor of 2 difference in the contact term with $(V_2+V_3)|_{VV}$ in the potential form. The $(V_2+V_3)|_{VV}$ potential we study has the form:

\begin{equation}
\label{gVgV_V23}
\begin{aligned}
&(V_2+V_3)|_{VV} = \\
&g_V^Xg_V^Y\frac{\hbar^3}{16\pi c m_Xm_Y}\left[ \boldsymbol{\sigma}_X \cdot \boldsymbol{\sigma }_Y  \left( \frac{1}{r^3} + \frac{1}{\lambda r^2} + \frac{1}{\lambda^2 r} - \frac{8 \pi}{3} \delta(\boldsymbol{r}) \right) \right. \\
&\left.-  \left( \boldsymbol{\sigma}_X \cdot \hat{\boldsymbol{r}} \right) \left( \boldsymbol{\sigma }_Y \cdot \hat{\boldsymbol{r}} \right)  \left( \frac{3}{r^3} + \frac{3}{\lambda r^2} + \frac{1}{\lambda^2 r} \right)  \right] e^{-{r}/{\lambda}} \, . 
\end{aligned}
\end{equation}
Here $\boldsymbol{\sigma}_{X}$ and $\boldsymbol{\sigma}_{Y}$ are vectors of Pauli matrices of the spins $\boldsymbol{s}_i=\hbar {\boldsymbol{\sigma}_i}/2$ of the two fermions, 
$r$ is the distance between particles $X$ and $Y$, $\hat{\boldsymbol{r}}$ is the unit position vector directed from particle $Y$ to particle $X$,
while $m_X$ and $m_Y$ denote the masses of fermions $X$ and $Y$, respectively. For the DT molecule, we use the equilibrium value $1.40108\,a_0$ \cite{huber_molecular_1979}.


Following the discussion in \citet{ledbetter_constraints_2013}, the $\delta$-function contribution to $V_{3}$ is neglected because the Coulomb repulsion of the two nuclei.  The measurements presented in Tab.\,\ref{table1} were carried out in the gas phase, so the internuclear vector $\hat{\boldsymbol{r}}$ suffers random reorientation due to collisions, leading to the averaging of Eq.\,\eqref{gVgV_V23}, see App.\,\ref{Spherical_Ave}. Similar powers in Eq.\,\eqref{gVgV_V23} are canceled on averaging, and also the $\delta$-function part of the potential does not contribute to any energy shift, thus the only remaining relevant term is $\propto 1/r$. 

This gives an effective exotic $J$-coupling $\Delta {J_3} \boldsymbol{I}_1 \boldsymbol{I}_2$, where $\boldsymbol{I}_{1,2}$ are the respective spins of the deuteron and tritium:
\begin{equation}
\label{J23}
\Delta J_{2+3}|_{VV}=-g_V^pg_V^N \frac{\hbar^3}{2\pi m_N^2 c} \frac{1}{3\lambda^2 r} e^{-{r}/{\lambda}}\,, 
\end{equation}
where $g_V^pg_V^N={g_V^Tg_V^D}/2$, with $g_V^T=g_V^p$ and $g_V^D=g_V^n+g_V^p \equiv 2\times g_V^N$. 
$m_N$ is the nucleon mass, assuming the neutron and proton masses are equal. For the duteron the spins of the proton and neutron add to a total nuclear spin 1 \cite{krane_introductory_1987}, so that $\langle\boldsymbol{I_2}\rangle={\hbar}\langle\boldsymbol{\sigma}_p\rangle={\hbar}\langle\boldsymbol{\sigma}_n\rangle$. 
For the triton, which is composed of one proton and two neutrons, given that the two neutrons pair up with a total spin 0, the overall spin of the tritium nucleus is determined by the remaining proton, which has a spin of {1}/{2} \cite{bloch_spin_1947}. Therefore, both DT and HD can provide us constraints on exotic $p$-$N$ interaction. 

\begin{figure}[t]
\centering
\includegraphics[width=0.5\textwidth]{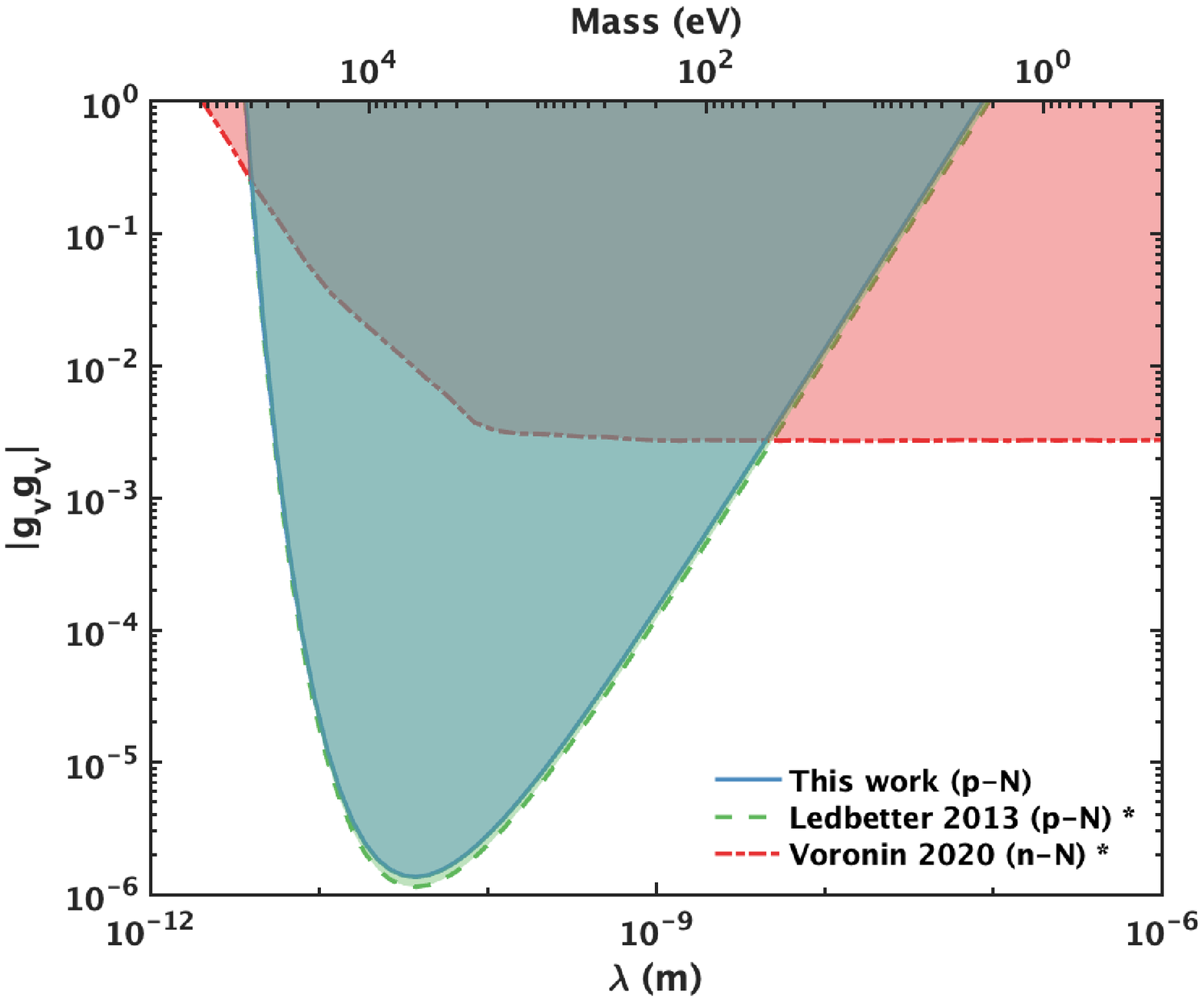}
\caption{Constraints, depicted by coloured regions, on the coupling constant product $g_V g_V$ as a function of the interaction range $\lambda$ shown on the bottom x-axis. The top x-axis represents the new spin-1 boson mass $M$. 
Constraints for exotic interaction between $p$-$N$ are obtained from comparison of measured and calculated $J$ constant for DT (solid blue curve, shaded light blue region). Others (labeled with *) are interpreted/translated in this work, based on data collected for HD \cite{ledbetter_constraints_2013} (dashed green curve, shaded light green region) and constraints for $n$-$N$ from \cite{voronin_constraint_2020} (dash-dotted red curve, shaded light red region), respectively.}
\label{fig:gVgV}
\end{figure}

The constraints for $g_Vg_V$ we obtained is show in Fig.\,\ref{fig:gVgV}. We also do the calculation for HD using the averaged theoretical and experimental results in \cite{ledbetter_constraints_2013} to obtain the constraints for $g_Vg_V$, based on Eq.\,\eqref{gVgV_V23} and \eqref{J23}. Note that there are recent experiments \cite{neronov_determination_2018} and theory \cite{puchalski_nuclear_2018} for HD, however, the implications of these will be discussed elsewhere. 
In addition, we present the $n$-$N$ constraints obtained from neutron-diffraction via potential $V_{4+5}$ \cite{voronin_constraint_2020}. We rescale their constraints for $g_Ag_A$ to $g_Vg_V$ \cite{cong_spin-dependent_2024,fadeev_neue_2018,dobrescu_spin-dependent_2006} by comparing $V_{4+5}|_{AA}$ and $V_{4+5}|_{VV}$: 
\begin{equation}
\label{gaga_V45}
\begin{aligned}
&V_{4+5}|_{AA}= \\
&g_A^Xg_A^Y \frac{\hbar^2}{16\pi c}\frac{m_X}{m_Y(m_X+m_Y)}\boldsymbol{\sigma}_X \cdot(\boldsymbol{v} \times \hat{\boldsymbol{r}}) \left (\frac{1}{r^2}+\frac{1}{\lambda r} \right ) e^{-r/\lambda} \, , \\
\end{aligned}
\end{equation}
and 
\begin{equation}
\label{gvgv_V45}
\begin{aligned}
&V_{4+5}|_{VV}= \\
&g_V^X g_V^Y \frac{\hbar^2}{16\pi c}\frac{2m_X+m_Y}{m_X(m_X+m_Y)}\boldsymbol{\sigma}_X \cdot(\boldsymbol{v} \times \hat{\boldsymbol{r}}) \left(\frac{1}{r^2}+\frac{1}{\lambda r}\right) e^{-{r}/{\lambda}} \, . 
\end{aligned}
\end{equation}

\subsection{Axial-vector/axial-vector interactions}
In the case of spin-1 axial vector (A) bosons, the exotic potential $V_{2}$ is \cite{fadeev_revisiting_2019}: 
\begin{equation}
\label{gaga_V2}
\begin{aligned}
V_{2}=-g_A^Xg_A^Y \frac{\hbar c}{4\pi}\boldsymbol{\sigma}_X\cdot\boldsymbol{\sigma}_Y\frac{1}{r}\,e^{-{r}/{\lambda}} \,, 
\end{aligned}
\end{equation} 
which gives
\begin{equation}
\label{J2}
\Delta J_{2}|_{AA}=-g_A^pg_A^N\frac{\hbar c}{\pi} \frac{1}{ r} e^{-{r}/{\lambda}}\,.
\end{equation}

\begin{figure}[t]
\centering
\includegraphics[width=0.5\textwidth]{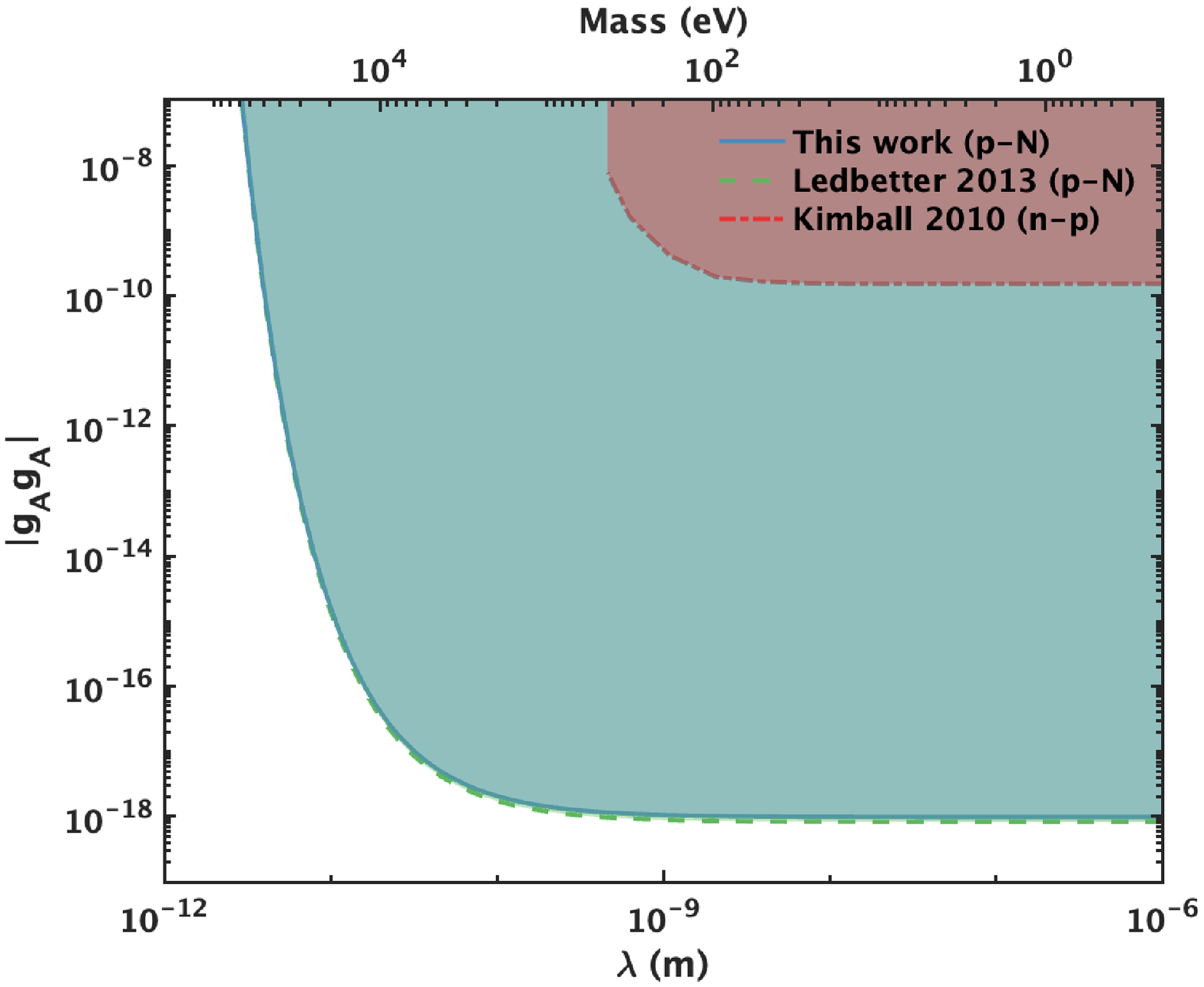}
\caption{Constraints, depicted by coloured regions, on the coupling constant product $g_A g_A$ as a function of the interaction range $\lambda$ shown on the bottom x-axis. The top x-axis represents the new spin-1 boson mass $M$. 
Constriants for exotic interaction between $p$-$N$ are obatined from comparaision of measured and calculated $J$ constant for DT (solid blue curve, shaded light blue region). The other one are constraints for $p$-$N$ from \cite{ledbetter_constraints_2013} (dashed green curve, shaded light green region) and constraints for $n$-$p$ from \cite{jackson_kimball_constraints_2010} (dash-dotted red curve, shaded light red region).}
\label{fig:gAgA}
\end{figure}

The constraint for $g_Ag_A$ obtained is shown in Fig.\,\ref{fig:gAgA}. The constraints for $p$-$N$ obtained from HD \cite{ledbetter_constraints_2013} and constraints for $n$-$p$ obtained by the study of cross sections for spin exchange between alkali metal atoms and noble gases \cite{jackson_kimball_constraints_2010} are presented for comparison.

The exotic potential $V_{3}|_{AA}$ is also rarely studied \cite{ fadeev_revisiting_2019, fadeev_pseudovector_2022}: 
\begin{equation}
\label{gaga_V3}
\begin{aligned}
V_{3}|_{AA}
&= - \hbar c {g_A^X g_A^Y \lambda^2}  \left[ \boldsymbol{\sigma}_X \cdot \boldsymbol{\sigma }_Y \left( \frac{1}{r^3} + \frac{1}{\lambda r^2} + \frac{4 \pi}{3} \delta(\boldsymbol{r}) \right) - \right. \\
&\left.  \left( \boldsymbol{\sigma}_X \cdot \hat{\boldsymbol{r}} \right) \left( \boldsymbol{\sigma }_Y \cdot \hat{\boldsymbol{r}} \right)  \left( \frac{3}{r^3} + \frac{3}{\lambda r^2} + \frac{1}{\lambda^2 r} \right)  \right] \frac{e^{-r/\lambda}}{4 \pi}\,,
\end{aligned}
\end{equation}
which gives
\begin{equation}
\label{J3AA}
\Delta J_{3}|_{AA}=g_A^pg_A^N\frac{\hbar c}{\pi} \frac{1}{3 r} e^{-{r}/{\lambda}}\,.
\end{equation}

Compared to Eq.\,\eqref{J2}, one can notice that the constraints for $g_Ag_A$ from $V_{3}$ is 3 times weaker than the one obtained from $V_{2}$. Thus it is not presented in Fig.\,\ref{fig:gAgA} for clarity. 

\subsection{Pseudoscalar/pseudoscalar interactions} \label{sec:gpgp}
The potential $V_3$ represents a dipole-dipole interaction generated by exchange of a pseudoscalar axion or axionlike particle (ALP) between fermions \cite{fadeev_revisiting_2019}: 
\begin{equation}\label{gpgp_V3}
\begin{aligned}
&V_{3}=\\
&-g_p^Xg_p^Y\frac{\hbar^3}{16\pi c}\frac{1}{m_Xm_Y}\left[\boldsymbol{\sigma}_X\cdot\boldsymbol{\sigma}_Y\cdot \left(\frac{1}{r^3}+\frac{1}{\lambda r^2}+\frac{4\pi}{3}\delta(r)^3\right)\right.\\
&\left.-(\boldsymbol{\sigma}_X\cdot \hat{\boldsymbol{r}})(\boldsymbol{\sigma}_Y\cdot \hat{\boldsymbol{r}})\left(\frac{3}{r^3}+\frac{3}{\lambda r^2}+\frac{1}{\lambda^2 r}\right)\right]e^{-{r}/{\lambda}}\,.
\end{aligned}
\end{equation}


Similar to the analysis in Sec.\,\ref{sec:gvgv}, we obtain:
\begin{equation}
\label{J3}
\Delta J_3|_{pp}=g_p^pg_p^N\frac{\hbar^3}{4\pi m_N^2 c} \frac{1}{3\lambda^2 r} e^{-{r}/{\lambda}}\,, 
\end{equation}

\begin{figure}[!htp]
\centering
\includegraphics[width=0.5\textwidth]{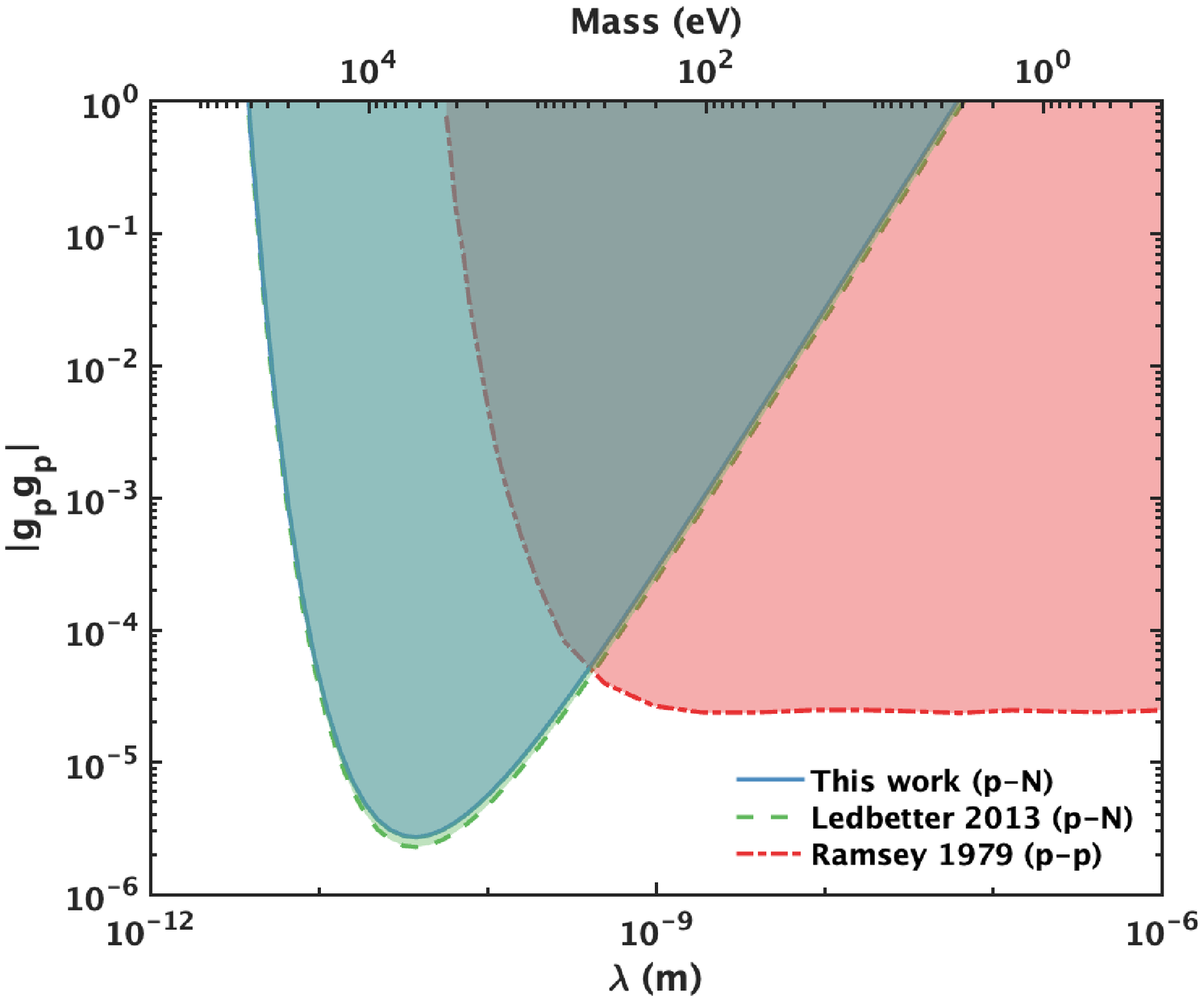}
\caption{Constraints, depicted by coloured regions, on the coupling constant product $g_p g_p$ as a function of the interaction range $\lambda$ shown on the bottom x-axis. 
The top x-axis represents the new spin-1 boson mass $M$. Constriants for exotic interaction between $p$-$N$ are obatined from comparaision of measured and calculated $J$ constant for DT (solid blue curve, shaded light blue region). Others are existing constraints for $p$-$N$ and $p$-$p$ from \cite{ledbetter_constraints_2013} (dashed green curve, shaded light green region) and \cite{ramsey_tensor_1979} (dashe-dotted red curve, shaded light red region), respectively.
}
\label{fig:gpgp}
\end{figure}

The constraints for $g_pg_p$ we obtained are shown in Fig.\,\ref{fig:gpgp}. For comparison, the limits obtained from Ramsey’s molecular-beam measurements \cite{ramsey_tensor_1979} of H$_2$ dipole-dipole interactions and the limits obtained from HD \cite{ledbetter_constraints_2013} are presented. There are also earlier and weaker constraints for exotic interactions between neutron and proton $n$-$p$ from \cite{jackson_kimball_constraints_2010}.

\section{Conclusion}
In summary, we used experimental and theoretical $J$-coupling values in tritium deuteride to constrain spin-dependent forces due to the exchange of exotic pseudoscalar, vector, and axial-vector particles. Stringent constraints are obtained for coupling strength combinations $g_pg_p$, $g_Vg_V$ and $g_Ag_A$ for exotic nucleon-nucleon interaction ($p$-$N$). Especially, following the latest theoretical framework \cite{fadeev_revisiting_2019,cong_spin-dependent_2024}, we first set constraints on $g_Vg_V$ by studying $(V_2+V_3)$. We noticed that the HD results \cite{ledbetter_constraints_2013} can also be interpreted as constraints on $g_Vg_V$. 

Further improvement relies on theoretical consideration of nonadiabatic and other corrections  \cite{puchalski_nuclear_2018} and more accurate experiments \cite{wilzewski_method_2017}. For example, if the $-0.10$ Hz correction estimated from nonadiabatic effects can be confirmed, the current constraints from DT could be improved by a factor of 5. Additionally, if the experimental uncertainty can simultaneously be reduced by a factor of 10, the overall improvement in the current constraints could be approximately a factor of 15. 

While some of our current results for the exotic couplings are similar to those of \citet{ledbetter_constraints_2013} (see Fig.\,\ref{fig:gpgp} and \ref{fig:gAgA}), it is important to emphasize that they are obtained from independent data for a different molecule.

\section*{Acknowledgment}
The authors thank Krzysztof Pachucki for valuable discussions.
This research was supported in part by the DFG Project ID 390831469: EXC 2118 (PRISMA+ Cluster of Excellence) and by the COST Action within the project COSMIC WISPers (Grant No. CA21106), the U.S. National Science Foundation grant PHYS-2110388, and the Munich Institute for Astro-, Particle and BioPhysics (MIAPbP), which is funded by the Deutsche Forschungsgemeinschaft (DFG, German Research Foundation) under Germany's Excellence Strategy – EXC-2094 – 390783311.

\section*{Appendix} 

\appendix 
\addcontentsline{toc}{section}{Appendix}

\section{Existing experiment and theoretical results for DT}\label{Existing_exp}

\begin{figure}[htbp]
\centering
\includegraphics[width=0.5\textwidth]{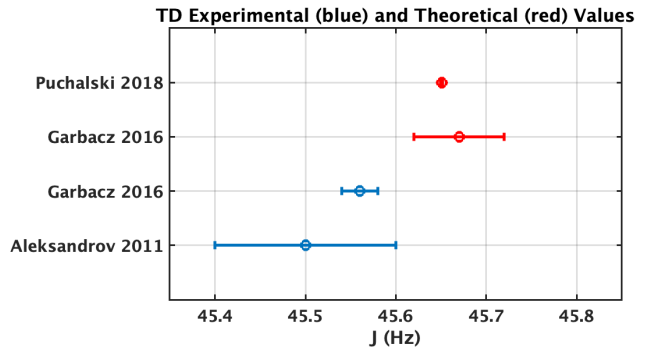}
\caption{Current experiment and theoretical study of the $J$-coupling constants of DT \cite{garbacz_indirect_2016,puchalski_nuclear_2018,aleksandrov_study_2011}. }
\label{fig:Existing-exp}
\end{figure}

Figure.\,\ref{fig:Existing-exp} presents all four existing experimental and theoretical results for the $J$-coupling constants of DT. Compare to earlier theoretical result $J=45.67(5)$ Hz from \cite{garbacz_indirect_2016} and experimental result $J=45.5(1)$ Hz from \cite{aleksandrov_study_2011}, the other two latest results have better uncertainty and are thus used in the main text to search for exotic spin-dependent interactions. 
Note that the earlier theoretical result \cite{garbacz_indirect_2016} did not involve non-adiabatic corrections.

\section{Spherical Averaging of Magnetic Dipole Interaction}\label{Spherical_Ave}

Consider two magnetic dipoles $\boldsymbol{\sigma}_X$ and $\boldsymbol{\sigma}_Y$. The spherical average of the interaction term $(\boldsymbol{\sigma}_X \cdot \hat{r})(\boldsymbol{\sigma}_Y \cdot \hat{r})$ over all orientations of $\hat{r}$ can be calculated as follows:

The unit vector in spherical coordinates is represented as:
\begin{equation}
\hat{r} = (\sin \theta \cos \phi, \sin \theta \sin \phi, \cos \theta)
\end{equation}

The dot products $(\boldsymbol{\sigma}_X \cdot \hat{r})$ and $(\boldsymbol{\sigma}_Y \cdot \hat{r})$ are:
\begin{equation}
(\boldsymbol{\sigma}_X \cdot \hat{r}) = \mu_{1x} \sin \theta \cos \phi + \mu_{1y} \sin \theta \sin \phi + \mu_{1z} \cos \theta
\end{equation}
\begin{equation}
(\boldsymbol{\sigma}_Y \cdot \hat{r}) = \mu_{2x} \sin \theta \cos \phi + \mu_{2y} \sin \theta \sin \phi + \mu_{2z} \cos \theta
\end{equation}

The product of these expressions is:
\begin{equation}
\begin{aligned}
(\boldsymbol{\sigma}_X \cdot \hat{r})(\boldsymbol{\sigma}_Y \cdot \hat{r}) &= (\mu_{1x} \sin \theta \cos \phi + \mu_{1y} \sin \theta \sin \phi + \mu_{1z} \cos \theta)\\
&(\mu_{2x} \sin \theta \cos \phi + \mu_{2y} \sin \theta \sin \phi + \mu_{2z} \cos \theta)
\end{aligned}
\end{equation}

Averaging over the sphere involves integrating this product over all directions and normalizing by the surface area of the sphere:
\begin{equation}
\langle (\boldsymbol{\sigma}_X \cdot \hat{r})(\boldsymbol{\sigma}_Y \cdot \hat{r}) \rangle = \frac{1}{4\pi} \int_0^{2\pi} \int_0^{\pi} (\boldsymbol{\sigma}_X \cdot \hat{r})(\boldsymbol{\sigma}_Y \cdot \hat{r}) \sin \theta \, d\theta \, d\phi
\end{equation}

Using the orthogonality relations of trigonometric functions over the integration interval:
\begin{equation}
\begin{aligned}
\int_0^{2\pi} \cos \phi \sin \phi \, d\phi = 0,\\ \quad \int_0^{2\pi} \cos^2 \phi \, d\phi = \int_0^{2\pi} \sin^2 \phi \, d\phi= \pi,\\ 
\int_0^{\pi} \sin^3 \theta \, d\theta = \frac{4}{3}, \quad \int_0^{\pi} \sin \theta \cos^2 \theta \, d\theta = \frac{2}{3}\,.
\end{aligned}
\end{equation}

These relations lead to the simplification and averaging:
\begin{equation}
\langle (\boldsymbol{\sigma}_X \cdot \hat{r})(\boldsymbol{\sigma}_Y \cdot \hat{r}) \rangle = \frac{1}{3} (\boldsymbol{\sigma}_X \cdot \boldsymbol{\sigma}_Y)
\end{equation}

In the DT system, after averaging the $(V_2+V_3)|_{VV}$ potential, it yielding an effective exotic J coupling $\Delta J_3 \boldsymbol{I}_1 \boldsymbol{I}_2 $ with 

\begin{equation}
\Delta J_3 = \frac{g_V^Tg_V^D}{2} \frac{1}{4\pi m_N^2} \frac{ 2 m^2 e^{-r/\lambda}}{3r}
\end{equation}

\bibliographystyle{apsrev4-2} 
\providecommand{\noopsort}[1]{}

\end{document}